\documentclass[aps,prd,showpacs,preprintnumbers,twocolumn]{revtex4-2}
\DeclareUnicodeCharacter{FFFD}{ }
\usepackage{amsmath,epsfig}
\usepackage{subfigure}
\usepackage{amssymb,amsfonts}
\usepackage{latexsym}
\usepackage{epsfig}
\usepackage{amssymb}
\usepackage{color}
\usepackage{bm}
\newcommand{\be}{\begin{eqnarray}}
\newcommand{\ee}{\end{eqnarray}}

\newcommand{\bea}{\begin{eqnarray}}
\newcommand{\eea}{\end{eqnarray}}

\def \and {{\textrm{and}}}

\begin{document}
\title{Zero-mode contribution and quantized first order phase transition \\
      in a droplet quark matter }
\author{Kun Xu$^{a,b}$ }
\thanks{xukun@mail.ihep.ac.cn, first author}
\author{Mei Huang$^{a}$}
\thanks{huangmei@ucas.ac.cn, corresponding author}
\affiliation{$^{a}$  School of Nuclear Science and Technology, University of Chinese Academy of Sciences, Beijing 100049, China}
\affiliation{$^{b}$ Institute of High Energy Physics, Chinese Academy of Sciences, Beijing 100049, P.R. China}

\begin{abstract}
The finite size effect on hadron physics and quark matter has attracted much interest for more than three decades, normally both the periodic (with zero-momentum mode) and the anti-periodic (without zero-momentum mode) spatial boundary condition are applied for fermions. By comparing the thermodynamical potential, it is found that if there is no other physical constraint, the droplet quark matter is always more stable when the periodic spatial boundary condition is applied, and the catalysis of chiral symmetry breaking is observed with the decrease of the system size, while the pions excited from the droplet vacuum keep as pseudo Nambu-Goldstone bosons. Furthermore, it is found that the zero-momentum mode contribution brings significant change of the chiral apparent phase transition in a droplet of cold dense quark matter: the 1st-order chiral apparent phase transition becomes quantized, i.e., the 1st-order apparent phase transition is completed in two steps, which is a brand-new quantum phenomena. It is expected that the catalysis of chiral symmetry breaking and the quantized 1st-order apparent phase transition are common features for fermionic systems with quantized momentum spectrum with zero-mode contribution, which also show up in quark matter under magnetic field.
\end{abstract}
\maketitle

\section{Introduction}

Size effect attracts wide interests in different physical systems. For example, in a recent PNAS article \cite{grapeplasma}, scientists realize that the most essential factor of making the grape plasma in microwave oven is the grape size, which is comparable with the typical microwave length, so that the grape can "trap" microwaves. Finite size effects on phase transitions were studied four decades ago \cite{Fisher:1972zza} and finite size effect in Quantum Chromodynamics (QCD) of hadron physics has attracted much interests for more than three decades \cite{Barber,Brezin:1985xx,Luscher:1985dn,Gasser:1987zq,Gasser:1987ah}, which is
important to extract hadron properties from numerical simulations on finite and discrete Euclidean space-time lattices. The study of finite size effect on quark matter and QCD phase structure becomes necessary and important \cite{Kiriyama:2006uh,Shao:2006gz,Palhares:2009tf,Braun:2005-2011,Zong-group,Almasi:2016zqf,Klein:2017shl}
due to the fact that the hot/dense matter created through heavy-ion collisions has finite volume with typical size of a nuclei.
Quark droplet has been also investigated in neutron stars \cite{Heiselberg:1992dx}. When consider the mixed phase of quark matter and nuclear matter, one has to solve the Wigner-Seitz cell structure, e.g., drop/bubble, rod/tube, and slab structure, and it is found that the size of the Wigner-Seitz cell can be as small as several fms as shown in \cite{Wu:2017xaz}.

As there is a typical length in the grape plasma, i.e., the microwave length, in QCD system the typical length is the pion Compton length $\lambda_{\pi}=1/m_{\pi}\sim 1.41 {\rm fm}$. When the system size is much larger than the pion Compton length $L >> \lambda_{\pi}$, hadron properties and phase transitions satisfy some finite size scaling (FSS) behavior \cite{Fisher:1972zza,Luscher:1985dn}. When the system size is comparable with or even smaller than the pion Compton length $L\sim \lambda_{\pi}$, the size effect becomes significant.

In the case of finite volume, the general method is to replace the momentum integral to momentum summation, i.e.
$\int \frac{d^3 p}{(2 \pi)^3}\rightarrow\frac{1}{V}\sum_{p}$, with $V$ the volume of system. For bosons, the most natural choice of boundary condition in the spatial direction is periodic, which implies that the momentum is summed from the exact zero-momentum mode ${\vec p} = 0$ in finite system. However, for fermions or quarks in finite system, its boundary condition (BC) in the spatial directions are not determined by a physical constraint, and one is free to choose either periodic (P) or anti-periodic (AP) boundary conditions. In lattice QCD simulations, the periodic boundary condition (PBC) is normally applied for fermions/quarks, and the anti-periodic boundary condition (APBC) is applied in most cases for fermions in the spatial direction in order to keep the so called permutation symmetry between the time and space directions \cite{Almasi:2016zqf,Klein:2017shl}, and also to be consistent with the results of volume dependent pion mass from chiral perturbation theory (ChPT) \cite{Luscher:1985dn}.

The boundary condition becomes important when the system size is comparable with the pion wavelength. Applying APBC and PBC to fermions induces two opposite results on the properties of QCD vacuum:  the APBC induces the chiral symmetry restoration, while the PBC induces the catalysis of chiral symmetry breaking in the vacuum, respectively. The catalysis of chiral symmetry breaking in small system with PBC including the zero-momentum mode contribution immediately reminds us the system of quark matter under strong magnetic field $B$ \cite{Gusynin:1994re,Miransky:2015ava}, where the summation of discrete energy is taken from the lowest Landau level, which is basically the zero-momentum mode. It is not difficult to understand the similarity between the two systems, if we recognize that for charged particle carrying charge $q$, the magnetic length $l$ is proportional to the inverse of the square root of the magnetic field, i.e. $ l\sim \frac{1}{\sqrt{|q|B}}$ \cite{Tong:2016kpv}, thus the stronger the magnetic field the smaller the magnetic length will be.

In this work, we carefully investigate quark matter in finite system with both the anti-periodic and periodic spatial boundary conditions applied for quarks  and analyze the two different physical results.  In Sec.II we compare the thermodynamical potential of the small size system applying PBC (with zero-mode) and APBC (without zero mode), respectively, and the lower thermodynamical potential determines the stable ground state of the small system. The results of catalysis of chiral symmetry breaking and pseudo NG pions are obtained in Sec.III, and then in Sec. IV we show the quantized 1st-order phase transition in cold droplet quark matter. At last we give summary and discussion. Besides, it is worth to point out that actual phase transitions are only possible at infinite volumes, thus, we use "apparent phase transition" instead for finite size systems.

\section{Droplet quark matter with PBC and APBC for quarks}

It is noticed that in this work, we only focus on discussing the boundary condition of fermions and neglect the finite size effect from gluon dynamics,
 therefore we can use the simplest four-fermion interacting Nambu--Jona-Lasinio (NJL) model.  The Lagrangian density of 2-flavor NJL model with
 interaction only in the scalar channel \cite{Klevansky:1992qe} is given by:
\be
\mathcal{L}=\bar{\psi}(i\gamma^\mu \partial_\mu -m)\psi+G[(\bar{\psi}\psi)^2+(\bar{\psi}\gamma^5 \vec{\tau}\psi)^2],
\ee
where $\psi=(u,d)^T$ is the doublet of the two light quark flavors $u$ and $d$ with the current mass $m=m_u=m_d$, and $\vec{\tau}=(\tau^1,\tau^2,\tau^3)$ the isospin Pauli matrix. Introducing the auxiliary scalar and pseudo-scalar fields $\sigma=-2G\langle\bar{\psi}\psi\rangle$ and
$\vec{\pi}=-2G\langle\bar{\psi}i\gamma_5\vec{\tau}\psi \rangle$, and only considering the scalar condensation in the vacuum, the effective potential of the system in the Pauli-Villars regularization scheme takes the form of:
 \be
 \Omega&=&\frac{\sigma^{2}}{4 G}-2 N_{c} N_{f} \int^{\infty}_{-\infty}\frac{d^3 p}{(2\pi)^3} {\large [}   \sum_{j=0}^{3} c_j \sqrt{E^2 + j\Lambda^2} \nonumber \\
  & & +T\ln(1+e^{-\frac{E+\mu}{T}})+T\ln(1+e^{-\frac{E-\mu}{T}}) {\large ]}  ,
 \ee
with $M=m+\sigma$ the constituent quark mass, and the dispersion relation $E=\sqrt{p^{2}+M^{2}}$, $T$ the temperature and $\mu$ the quark chemical potential, respectively. Model parameters are fixed as $c_0=1,\quad c_1=-3, \quad c_2=3, \quad c_3=-1$ and $\Lambda=782.37 \text{MeV}$  and $G=6.197/\Lambda^2 $ by fitting pion decay constant $f_{\pi} =93\text{MeV}$, and quark constitute mass $M=330\text{MeV}$.

\begin{figure}
\centering
\includegraphics[width=0.45\textwidth]{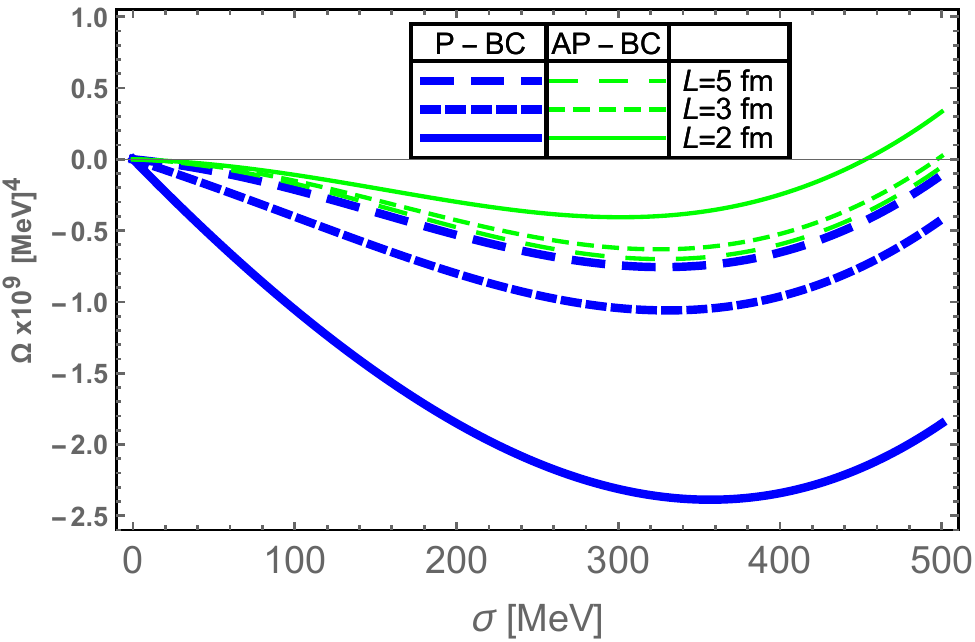}
\caption{The effective potential of the small system for three different sizes $L=5,3,2 ~ fm$ as a function of the chiral condensate $\sigma$ at $T=0,\mu=0$ by applying the periodic (blue lines) and anti-periodic (green lines) spatial boundary conditions for quarks, respectively. }
\label{fig:omega}
\end{figure}

\begin{figure}
\centering
\includegraphics[width=0.45\textwidth]{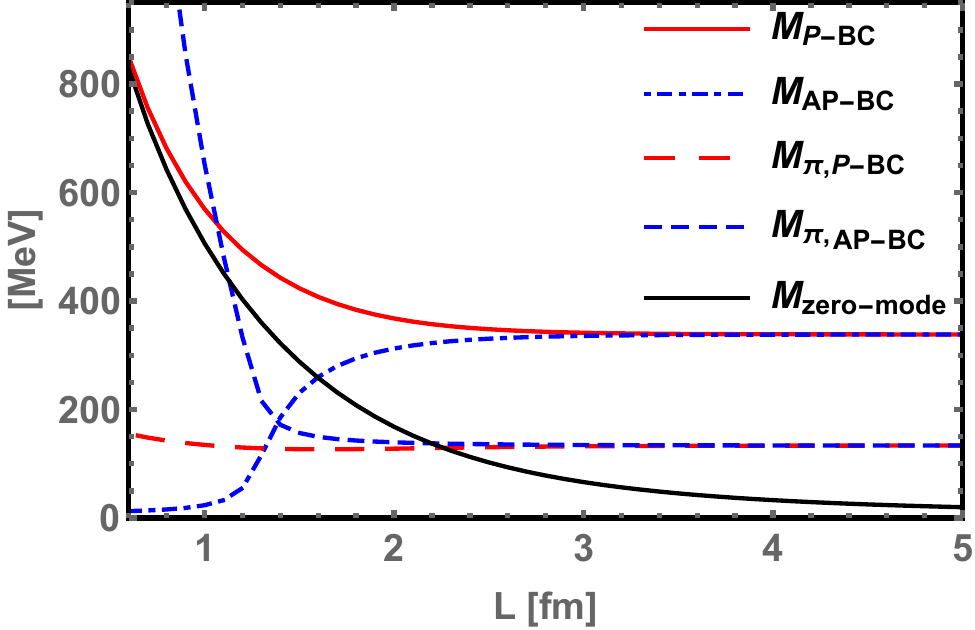}
\caption{The constituent quark mass and pion mass as a function of the system size $L$ at $T=0,\mu=0$ by applying the P-BC and AP-BC for quarks, respectively. $M_{\text{zero-mode}}$  is obtained with only zero-mode contribution. }
\label{fig:M-P-AP-PV}
\end{figure}

Putting quark matter in a cubic box with finite length $L$, the momentum integral is replaced by summation of the discrete momentum
$\int \frac{d^3 \vec{p}}{(2 \pi)^3}\rightarrow\frac{1}{V}\sum_{\vec{p}}$ with $V=L^3$ the volume of the system. There is no strict rule to rule out
either PBC ${\vec p}^{2}=(\frac{2\pi}{L})^{2}\sum_{i=x,y,z}n_{i}^{2}$ or APBC ${\vec p}^{2}=(\frac{2\pi}{L})^{2}\sum_{i=x,y,z}(n_{i}+\frac{1}{2})^{2}$ with $n_i=0,1,2,...$ for fermion momentum summation, therefore both spatial boundary conditions have been applied for fermions for several decades in literatures \cite{Kiriyama:2006uh,Shao:2006gz,Palhares:2009tf,Braun:2005-2011,Zong-group,Almasi:2016zqf,Klein:2017shl}.

It is well-known the ground state of the system is determined by the effective potential. We show the effective potential of the finite size system by applying PBC and APBC boundary condition, respectively in Fig.\ref{fig:omega}. At $T=0,\mu=0$, it is observed that when applying the PBC for quarks, the effective potential becomes lower with decreasing size, while when applying the APBC for quarks, the effective potential becomes higher with decreasing size. This indicates that if there is no other physical constraint, the quark-droplet prefers the PBC for quarks, in which the zero-momentum mode is taken into account.

From the effective potential in Fig. \ref{fig:omega}, we can read the chiral condensation in the vacuum, and the corresponding constituent quark mass and pion mass are shown in Fig. \ref{fig:M-P-AP-PV} by applying the PBC and APBC for quarks, respectively. We can read that when applying the PBC for quarks, with the decreasing of the system size, the chiral condensate enhances while pion mass keeps as a constant. This is the familiar phenomenon of catalysis of chiral symmetry breaking also observed in quark matter under strong magnetic fields, where only neutral pion keeps as pseudo NG boson. On the
other hand, if APPC is applied for quarks, it is found that the chiral symmetry is restored in small system and pions become heavier in the vacuum.

The main difference between the PBC and APBC is whether to take into account the zero-momentum mode contribution. For discrete momentum, the gap between the zero-mode and the first-mode is $\frac{2\pi}{L}$, therefore, when the system becomes small enough, the zero-momentum mode contribution becomes dominant which can be seen from the constituent quark mass with only zero-mode contribution, as shown in Fig. \ref{fig:M-P-AP-PV}. This is also the same as the case of strong magnetic field where the lowest Landau level contribution dominates. Remember that for charged particle carrying charge $q$, the magnetic length $l$ is proportional to the inverse of the square root of magnetic field, i.e. $ l\sim \frac{1}{\sqrt{|q| B}}$, thus putting charged particles under the strong magnetic field in some sense is similar to put these particles in an elongated cylinder with small radius $l$.

\begin{figure}
\centering
\subfigure[MIT bag model] { \label{fig:a}
\includegraphics[width=0.13\textwidth]{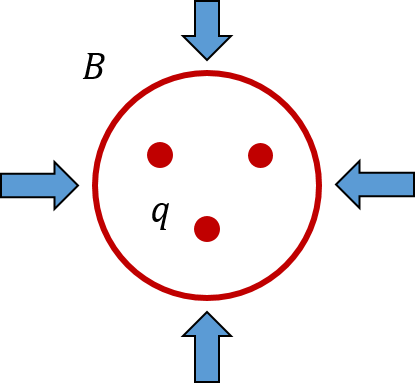}
}
\subfigure[Quark-droplet with PBC] { \label{fig:b}
\includegraphics[width=0.13\textwidth]{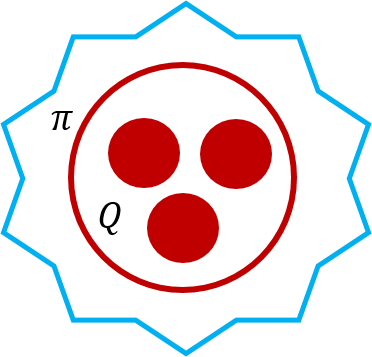}
}
\subfigure[Quark-droplet with APBC] { \label{fig:c}
\includegraphics[width=0.13\textwidth]{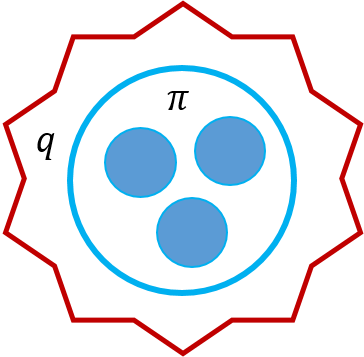}
}
\caption{The schematic picture for three scenarios of droplet quark matter: (a) The MIT bag model consists of free quarks q with the bag constant B putting by hand at the surface of the bag. (b) The quark droplet with periodic boundary condition consists of massive quarks Q with catalysis of chiral symmetry breaking and the pseudo Nambu-Goldstone pions' cloud. (c) The quark droplet with anti-periodic boundary condition consists of heavy pions inside the bag and light quarks' cloud. It is noticed that here the numbers and colors of quarks are showing in a schematic way. }
\label{fig:MITbag-Droplet-P-AP}
\end{figure}

Furthermore, it is very interesting to notice that the difference of the thermodynamical potential or the pressure difference between L=2 fm and L=$\infty$ is in the order of $[200 {\rm MeV}]^4$, which is  the typical value of bag constant in the MIT bag model \cite{MITbagmodel}. If we consider the spherical shape droplet quark matter, we will obtain qualitatively the same results as that for cubic droplet quark matter shown in Fig.1 and Fig.2 when PBC and APBC boundary conditions applied for quarks, respectively. In Fig. \ref{fig:MITbag-Droplet-P-AP}, we show three scenarios of quark droplet: (a) MIT bag model, (b) quark droplet with PBC, and (c) quark droplet with APBC. The MIT bag model consists of free quarks q with the bag constant B putting by hand at the surface of the bag, and the quark-droplet with PBC consists of massive quarks Q with catalysis of chiral symmetry breaking and the pseudo Nambu-Goldstone pion cloud. However, for the quark-droplet with APBC, the quarks restore chiral symmetry and the quark mass becomes very small, instead the pions become very heavy, therefore we can imagine that the heavy pions stay inside the bag with light quark cloud. We can see that the quark-droplet with PBC is very similar to the scenario of the MIT bag model, except that we can have spontaneous chiral symmetry breaking as well as Nambu-Goldstone pions in the quark-droplet with PBC. Therefore, the quark-droplet with PBC looks like a bag model with dynamical quarks, on the other hand, the quark-droplet with APBC is a bag with heavy pions. Of course, in this paper we are not intending to derive the MIT bag model, we just want to show that the droplet of quark matter under periodic boundary condition inducing catalysis of chiral symmetry breaking and Numbu-Goldstone pions simultaneously is a reasonable result in the vacuum. In the future, it might be interesting to re-investigate the bag model from the point of view of droplet-quark matter.

\section{Quantized 1st order apparent phase transition in cold droplet quark matter}

In last section, we showed the ground state of the small system favors applying the PBC for quarks, and the zero-mode contribution becomes dominant in small system. Now we investigate the chiral apparent phase transition at high temperature and/or baryon density with changing of the finite size. For infinite volume system, the NJL model predicts a critical end point for chiral phase transition in the $(T,\mu_B)$ plane located at $(T^E=48 {\rm MeV},\mu_B^E=994 {\rm MeV})$ with $\mu_B=3\mu_q$ the baryon chemical potential. When the size decreases from $L=\infty$ to $L=5 {\rm fm}$, which is much larger than the pion wavelength $\lambda_{\pi}$, there is almost no changing for the properties of the system.

\begin{figure}
\centering
\includegraphics[width=0.45\textwidth]{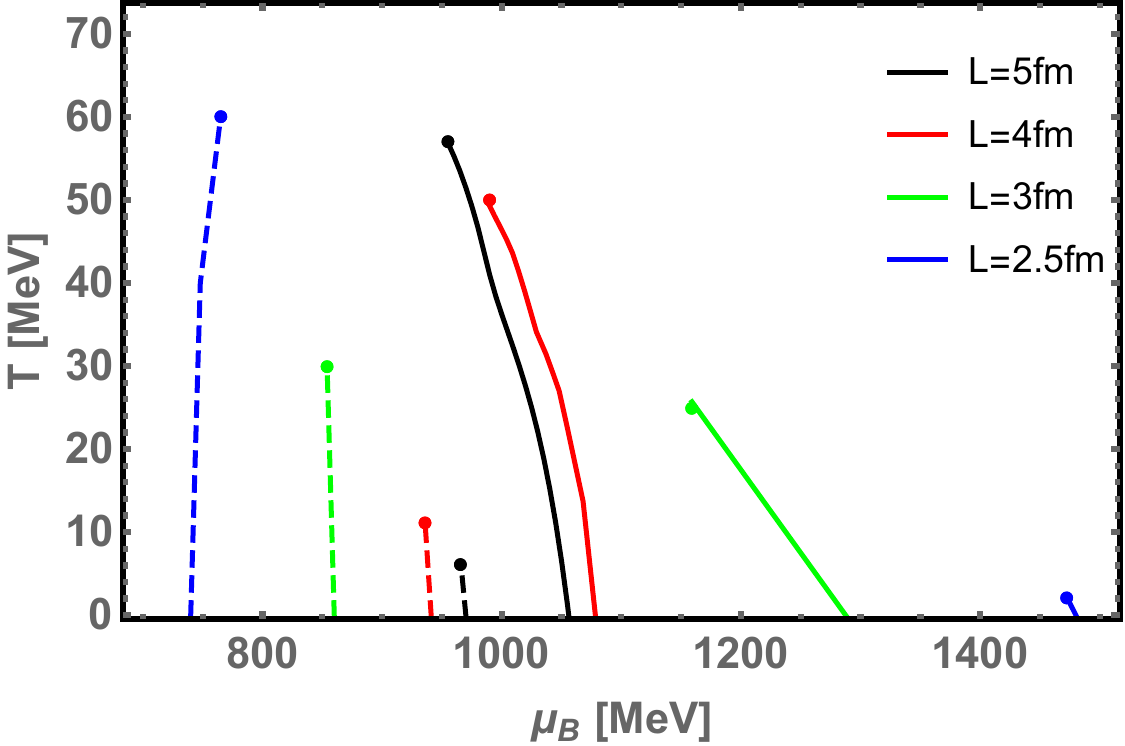}\hspace{20pt}
\caption{Two branches of first order apparent phase transitions and two sets of apparent critical end points in the $(T,\mu_B)$ plane for different sizes $L=5,4,3,2.5,2 fm$. The dashed lines and solid lines correspond to APT1 and APT2, respectively.}
\label{fig:ceps}
\end{figure}

When the size further decreases, it is found that the properties of the system start to change dramatically. As shown in Fig.\ref{fig:ceps},  in certain region of the chemical potential, there are two branches of  1st-order apparent phase transition(APT) showing up and two apparent CEPs appear on the $(T,\mu_B)$ plane for different sizes $L=5,4,3,2.5,2 fm$ with the dashed lines corresponding to the branch of apparent phase transition at lower chemical potential marked as APT1 and the solid lines corresponding to the branch of apparent phase transition at larger chemical potential marked as APT2, respectively. The apparent CEPs corresponding to APT1 and APT2 are marked as "CEP1"  and "CEP2" correspondingly. From Fig.\ref{fig:ceps}, it is observed that when the size decreases, CEP1 and CEP2 of these two branches have opposite behaviors: CEP2 moves to higher chemical potential and lower temperature, and CEP1 moves to lower chemical potential and higher temperature. When the size further decreases, the APT2 and CEP2 disappears and only the APT1 and CEP1 shows up. These results were also observed in Ref.\cite{Almasi:2016zqf}.

\begin{figure}
\centering
\includegraphics[width=0.45\textwidth]{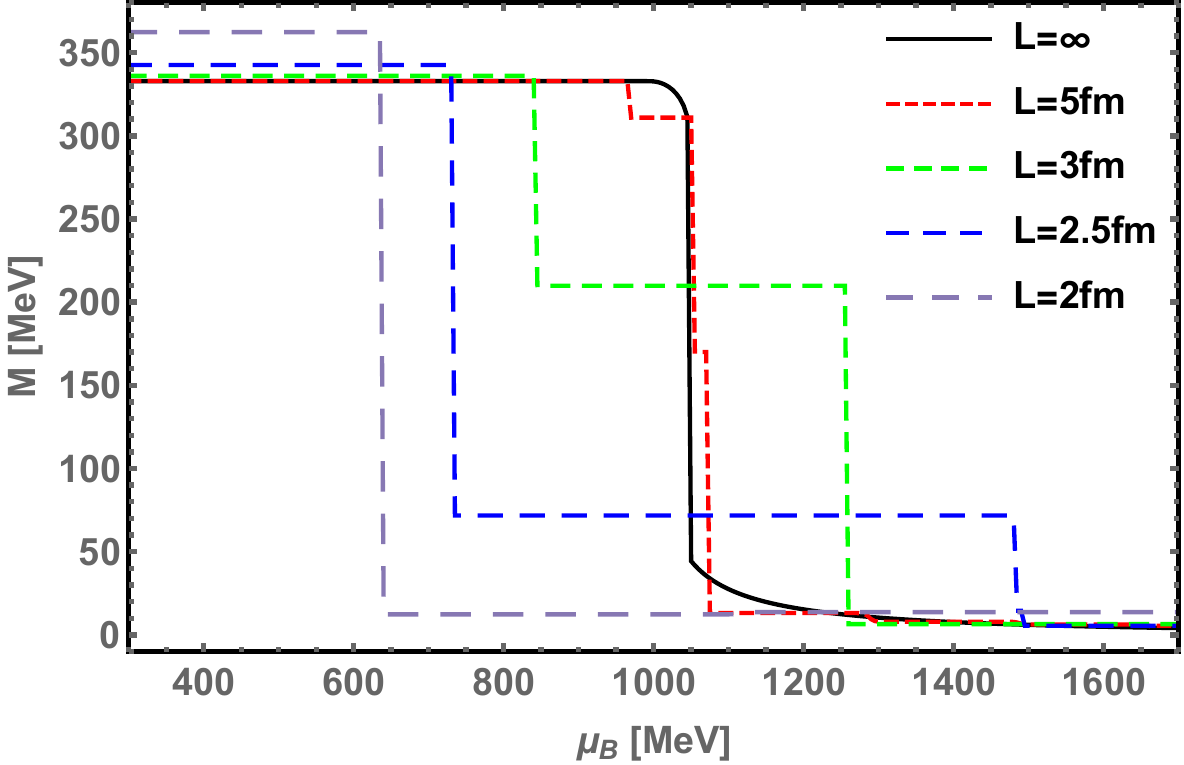}
\caption{The constituent quark mass as a function of the baryon chemical potential $\mu_B=3\mu $ with different sizes of the system at zero temperature.}
\label{fig:m-muB-L-PV}
\end{figure}

The showing up of two branches of 1st-order apparent phase transition can be clearly explained through Fig. \ref{fig:m-muB-L-PV}, which shows the constituent quark mass as a function of the baryon chemical potential $\mu_B=3\mu $ with different sizes of the system at zero temperature. In the region of $2~fm<L<5~fm$, the constituent quark mass jumps twice with the increase of the chemical potential. This multi-jump structure is caused by the zero-mode contribution. With the decrease of the system size, the zero-momentum contribution becomes more and more important. The first jump of the quark mass happens at lower critical chemical potential and the magnitude of the jump becomes bigger and bigger, and the second jump of the quark mass appears at higher chemical potential and its magnitude becomes smaller and smaller. Eventually at small enough size when the zero-mode contribution dominates only the first jump shows up. Correspondingly, the location of the first CEP, i.e., CEP1 shifts to lower  critical chemical potential but higher critical temperature and the location of the second CEP, i.e., CEP2 moves to higher and higher critical chemical potential but lower critical temperature.

We call this multi-jump 1st-order apparent phase transition as quantized 1st-order apparent phase transition, where each 1st-order apparent phase transition is finished in several steps. We want to emphasize that such quantized 1st-order apparent phase transition induced by quantized momentum is a brand-new quantum phenomena.

To show more clearly how these two branches of apparent phase transitions evolves with the system size, we show the 3-dimension (3D) plot of the kurtosis of baryon number fluctuations $\kappa \sigma^{2}$ in the $(T,\mu_B)$ plane. The ratio of the fourth to the second order cumulant of quark number fluctuations $\kappa \sigma^{2}=\frac{c_{4}}{c_{2}}$ with $c_{n}=VT^{3}\frac{\partial^{n}}{\partial(\mu/T)^{n}}(\frac{p}{T^{4}})$ is used as a measurement to locate the apparent CEP in beam-energy scan at RHIC experiment \cite{Luo:2017faz}. At $L=\infty$, there is only one typical 1st-order phase boundary. It is observed clearly that when the system size decreases, two branches of 1st-order apparent phase transition show up on the $(T,\mu_B)$ plane, one branch moves to higher chemical potential and eventually disappears, and another branch shifts to lower chemical potential region and then becomes dominant.

\begin{figure}
\centering
\includegraphics[width=0.45\textwidth]{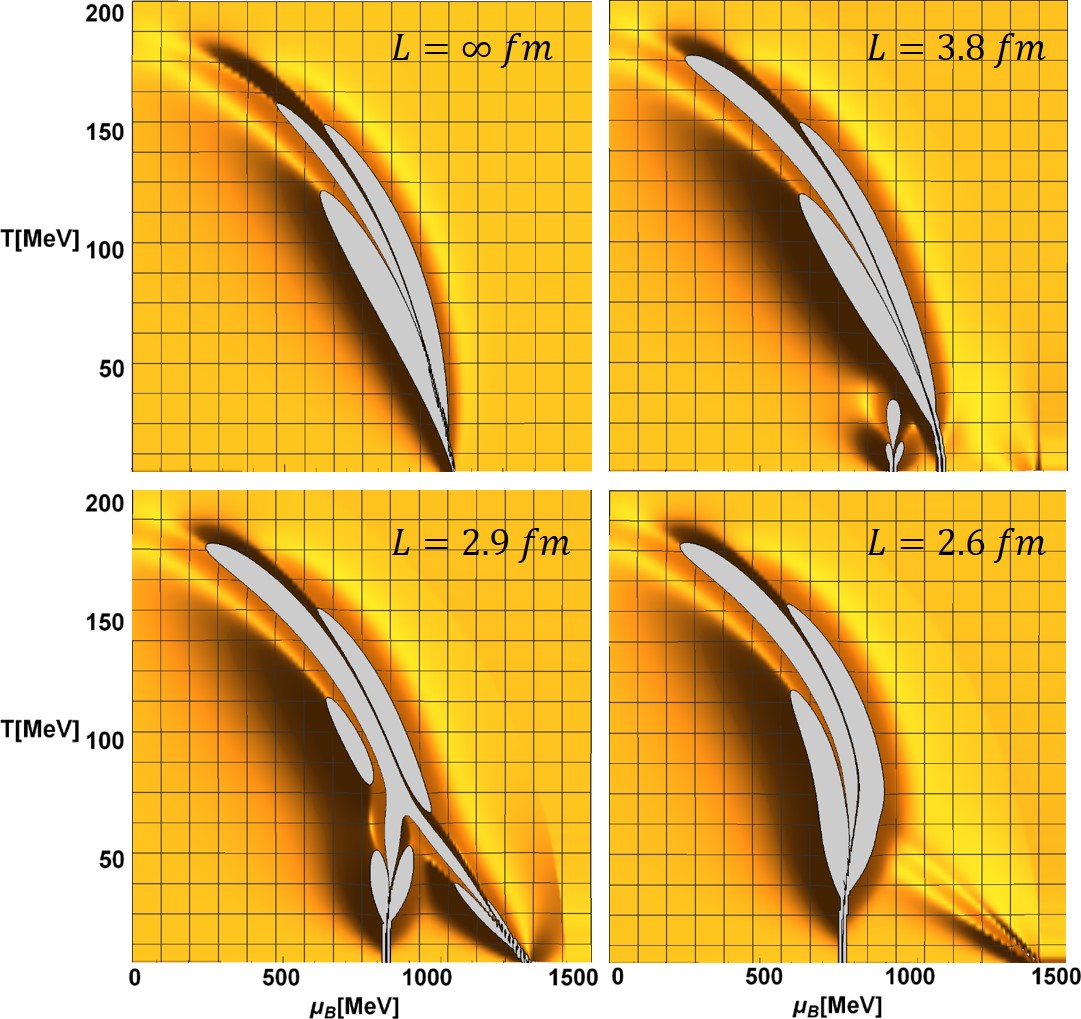}
\caption{The 3D plot of $\kappa \sigma^2$ in $(T,\mu_{B})$ plane with different sizes of the system.}
\label{fig:4-1}
\end{figure}

\section{Analyzing the appearance of quantized 1st-order apparent phase transition}

In the following, we analyze why the multi-jump structure of the 1st-order apparent phase transition can show up in certain small size system, which was shown in our proceeding paper Ref.\cite{Xu:2019kzy}. For this purpose, it can be much more transparent and simpler if we use the NJL model in the hard-cutoff regularization scheme, in which we can have the same qualitative results as that in the Pauli-Villars regularization scheme. By taking the hard-cutoff regularization for the momentum summation, the thermodynamical potential has the form of :
 \begin{eqnarray}
 & & \Omega_{\Lambda}=\frac{(M-m_{0})^{2}}{4 G}-\frac{2 N_{c} N_{f}}{V} \sum_{\vec{p}} \nonumber \\
 & & \left\lbrace E +T\ln(1+e^{-\frac{E+\mu}{T}})+T\ln(1+e^{-\frac{E-\mu}{T}})\right\rbrace .
 \end{eqnarray}
Here the momentum should be smaller than the hard-cutoff $\Lambda$:
$\Lambda^{2}>p^{2}=n(\frac{2\pi}{L})^{2}$.

In the following we only consider the system at zero temperature $T=0$, and now the gap equation:
\begin{equation}
 \frac{M-m}{2G} = \frac{2 N_{c} N_{f}}{V}\sum_{\vec{p}}\frac{M}{E}[1-\theta(\mu-E)] 
 \label{eq-gap:gapequ}
\end{equation}
here $\theta(x)$ is the step function. In the case of small enough size so that $2\pi/L>\Lambda$ with only the zero-momentum mode $n=0$ contributes to the momentum summation, the gap equation takes the form of:
\begin{equation}
\frac{M-m}{2G}=\frac{2 N_{c} N_{f}}{V}[1-\theta(\mu-E_0)],
\label{eq-gap:zeromode}
\end{equation}
and now $M=E_0=\sqrt{M^{2}+0(2\pi/L)^{2}}$. The solution to this equation is simple and given below:
\begin{equation}
M-m=\left\{
\begin{array}{lr}
	\frac{4GN_c N_f}{V}, \quad\mu<\mu^{c}  & \\
	0\quad\quad ,\quad \quad \mu>\mu^{c} &
\end{array}
\right.,
\end{equation}
with $\mu^{c}$ the critical quark chemical potential for chiral apparent phase transition at zero temperature. Obviously, above analysis is consistent with the line $L=2fm$ in Fig.\ref{fig:m-muB-L-PV}.

Then we consider a little bit bigger size so that both $n=0$ mode and $n=1$ mode can contribute to the momentum summation, and the gap equation takes the form of:
\begin{eqnarray}
\frac{M-m}{2G}&=&\frac{2 N_{c} N_{f}}{V}[1-\theta(\mu-E_{0})] \nonumber \\
& & +6\frac{2 N_{c} N_{f}}{V}\frac{M}{E_{1}}[1-\theta(\mu-E_{1})],
\label{eq-gap:firstmode}
\end{eqnarray}
where $E_1=\sqrt{M^{2}+(2\pi/L)^{2}}$. The second term at the right-hand side is from the contribution of the first-mode and 6 is the degeneracy number of the 1st-mode. The solution to this equation is not as straightforward as Eq.(\ref{eq-gap:zeromode}), but can be calculated numerically.

\begin{figure}
\centering
\includegraphics[width=0.45\textwidth]{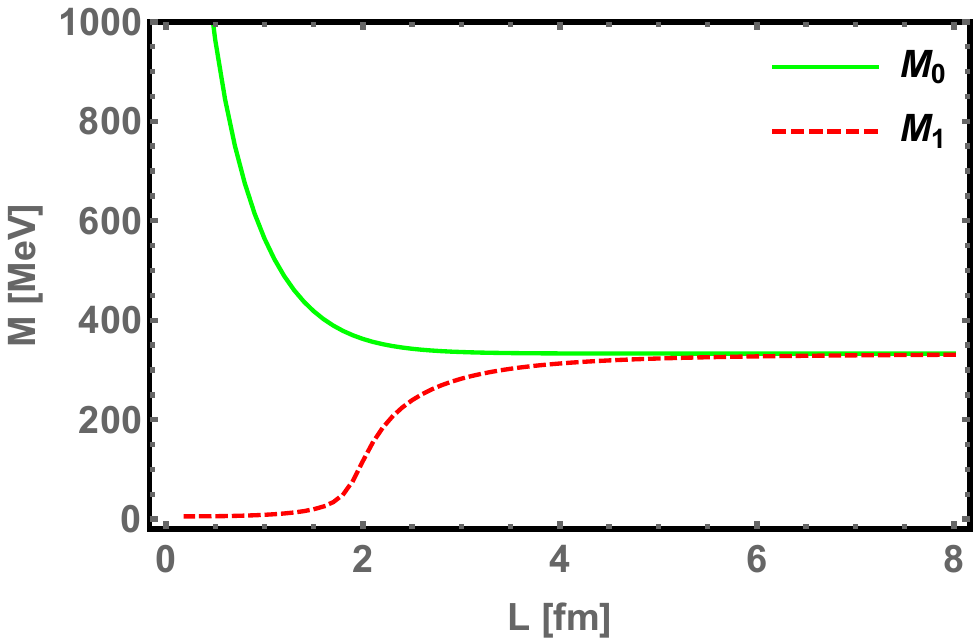}\hspace{20pt}
\caption{The constituent quark mass $M_0$ with zero-mode contribution and $M_1$ without zero-mode contributions as a function of the size $L$ at zero temperature and zero chemical potential. Here we take the chiral limit $m=0$.}
	\label{fig:m0-m1-L}
\end{figure}

We consider the chiral limit case with $m=0$. To show how zero mode influence constituent quark mass,  in Fig. \ref{fig:m0-m1-L}, we plot the constituent quark mass as a function of size  $L$ at zero temperature and zero chemical potential, where $M_0$ and $M_1$ are obtained by solving gap equation Eq.(\ref{eq-gap:gapequ}) with and without zero mode contribution, respectively. It can be seen clearly that: 1) At large size when $L>5  fm$, $M_0$ and $M_1$ are almost equivalent, which indicates that the contribution from the zero mode can be ignored when the size is large enough; 2) When the size decreases and becomes smaller than $5 fm$, $M_0$ and $M_1$ show completely opposite behaviors: $M_0$ rises quickly and goes to divergence with the decrease of the size, while $M_1$ decreases smoothly to zero at very small size. In general, more step functions can appear in larger sizes therefore more jumps are expected. However, higher modes would not contribute a lot. For example, the second term in the right-hand side of Eq.(\ref{eq-gap:firstmode}) is quite small comparing with the first term. This can be also verified from Fig.\ref{fig:m-muB-L-PV} that the magnitude of the second jump PT2 at $L=2.5fm$ is smaller than that at $L=3fm$.

With the help of $M_0$ and $M_1$, we can now understand the quantized first-order apparent phase transition in the size region of $2 fm<L<5 fm$. There are two apparent phase transitions APT1 and APT2 as shown in Figs.\ref{fig:ceps} and \ref{fig:m-muB-L-PV}, where APT1 can be understood as the jump between the chiral symmetry breaking phase with quark mass $M_0$ and the chiral partially "restoring" phase with quark mass $M_1$,  and APT2 can be understood as the jump between the phase with quark mass $M_1$ and the chiral symmetry fully restoration phase with $M=0$.  At large size when $L>5 fm$, we can see that $M_0 =M_1$, which indicates that the first branch phase transition PT1 vanishes. At small size when $L<2fm$ where $M_1=0$, the second branch apparent phase transition APT2 vanishes. Therefore, there is only one jump of the 1st-order apparent phase transition at both cases with large size $L>5 fm$ and small size $L<2 fm$, but there appears two jumps for the 1st-order apparent phase transition in the region of $2fm <L< 5 fm$. This is exactly what have obtained in Fig.\ref{fig:ceps} and Fig.\ref{fig:m-muB-L-PV}.

From above analysis, we can expect that  the quantized 1st-order apparent phase transition are common features for fermionic systems with quantized momentum spectrum with zero-mode contribution.

\section{Catalysis of chiral condensate and quantized 1st-order phase transition under strong magnetic field}
\begin{figure}	
	\centering
	\includegraphics[width=200pt]{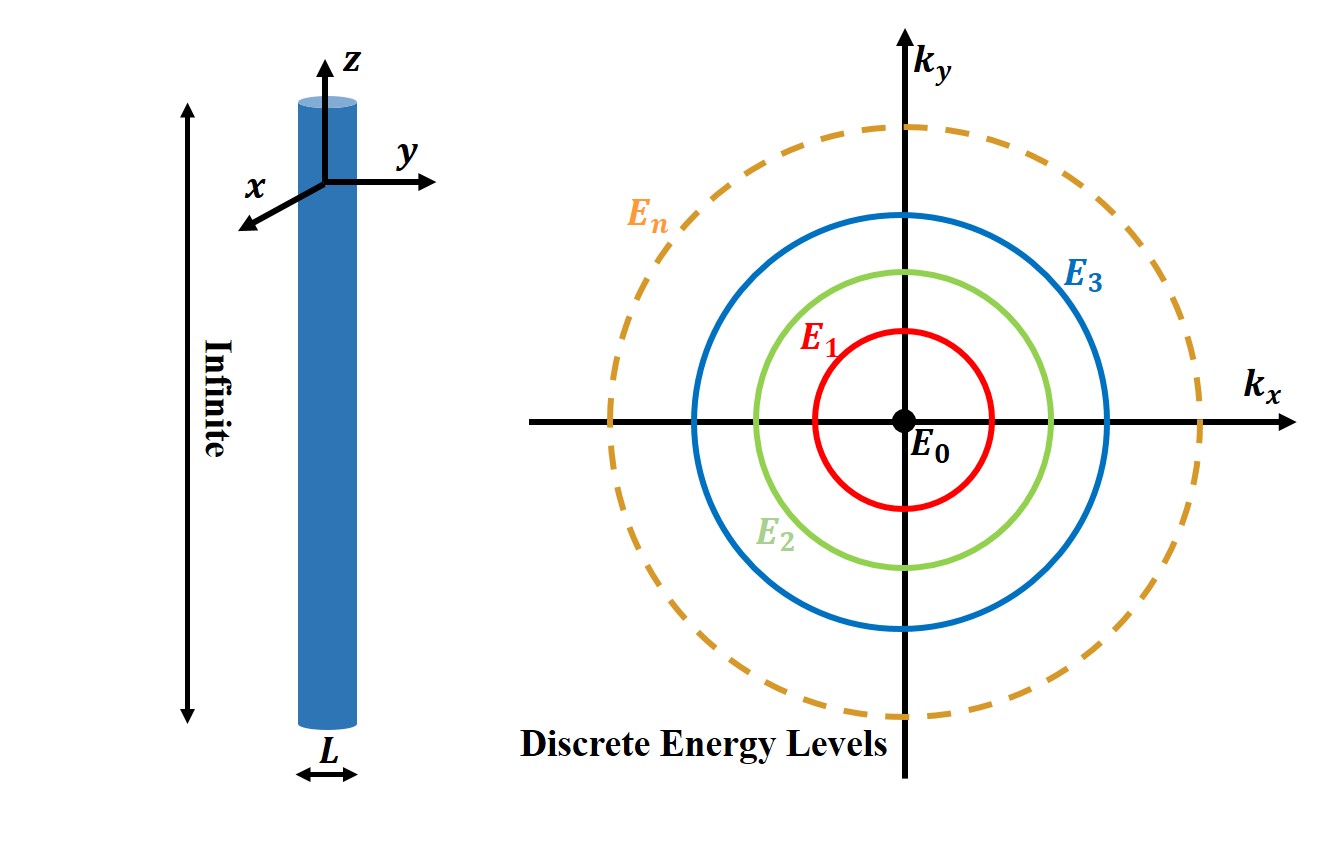}
	\caption{The cylinder with finite size at $x-y$ plane and the discrete-Energy-Levels.}\label{fig:Discrete-Energy-Levels}
\end{figure}

Now consider a cylinder system with infinite size along $z$-axis but finite size in $x-y$ plane as shown in Fig. \ref{fig:Discrete-Energy-Levels}, then momentum integral in the thermodynamic potential calculation is replaced by momentum integral in $z$-axis and momentum summation in  $x-y$ plane:
 \begin{eqnarray}
& & \Omega=\frac{(M-m_{0})^{2}}{4 G}-\frac{2 N_{c} N_{f}}{S} \sum_{\vec{p}_{\perp}}\int_{-\infty}^{\infty}\frac{dp_z}{2\pi} \nonumber \\
& & \left\lbrace E +T\ln(1+e^{-\frac{E+\mu}{T}})+T\ln(1+e^{-\frac{E-\mu}{T}})\right\rbrace ,
\end{eqnarray}
where S is the area of cylinder and $p_{\perp}$ the discrete momentum in $x-y$ plane. $p_{\perp}$ can be obtained by applying boundary conditions on the surface in radius direction, for example, a boundary condition that requires wave function vanishes on the surface. By applying different boundary conditions, different discrete momenta can be obtained and determines that whether zero-momentum mode is included. And in above equations, we sum all momenta in $x-y$ plane start from $p_{\perp}=0$.

As we mentioned in the Introduction, the catalysis of chiral symmetry breaking in small system with PBC including the zero-momentum mode contribution is similar with the system of quark matter under strong magnetic field $B$. It is not difficult to understand the similarity between the two systems, if we recognize that for particle with charge $q$, the magnetic length $l$ is proportional to the inverse of the square root of the magnetic field, i.e. $ l\sim \frac{1}{\sqrt{|q|B}}$ \cite{Tong:2016kpv}, which is similar to a cylinder system as shown in Fig. \ref{fig:Discrete-Energy-Levels}, and the stronger the magnetic field the smaller the magnetic length will be. In the presence of magnetic field, the momentum integral is replaced by discrete momentum summation in the plane perpendicular to magnetic field as shown in Fig. \ref{fig:Discrete-Energy-Levels}. Therefore, we expect that the quantized 1st-order phase transition can also show up in cold quark matter under magnetic field.

The thermodynamic potential of quark matter in the presence of magnetic field takes the form as given in \cite{Chao:2013qpa}:
\begin{eqnarray}
       \Omega  &=& \frac{\sigma^2}{4G}-N_c \sum_{f=u,d}\frac{|q_f B|}{2\pi} \sum_{s,k}\alpha_{s,k}\int_{-\infty}^{\infty}\frac{dp_z}{2\pi}\omega_{k}(p)  \nonumber \\
			& & -T N_c \sum_{f=u,d}\frac{|q_f B|}{2\pi}\sum_{s,k}\alpha_{s,k}\int_{-\infty}^{\infty}\frac{dp_z}{2\pi} \nonumber \\
    & &[\ln(1+e^{-\beta (\omega_{k}+\mu)})+\ln(1+e^{-\beta  (\omega_{k}-\mu)})].
\end{eqnarray}
Here $\omega_{k}=\sqrt{(\sigma+m_0)^2 +p_z^2+2|q_f B|k}$ is the dispersion relationship and $k=0,1,2,...$ a non-negative integer number labeling the Landau levels, and the spin degeneracy factor $\alpha_{s,k}$ is given below:
\begin{equation}
\alpha_{s,k}=
	\begin{cases}
	 \delta_{s,+1} & \text{for}\quad k=0, \quad qB>0, \\
	 \delta_{s,-1} & \text{for}\quad k=0, \quad qB<0, \\
	 1             & \text{for}\quad k\neq 0 \quad,
	\end{cases}
\end{equation}
with $s=\pm$ the spin factor. Here we take the magnetic field $B$ along z-axis and only consider $u$ and $d$ quarks. Similar to the case without magnetic field, the ground state of the system can be determined by solving the gap equation $\partial \Omega /\partial \sigma =0$.

\begin{figure}
	\centering
	\includegraphics[width=200pt]{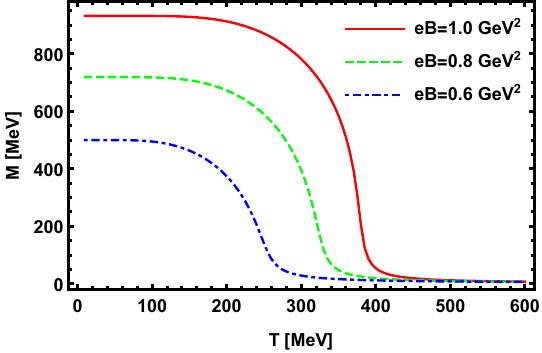}\\
    \includegraphics[width=200pt]{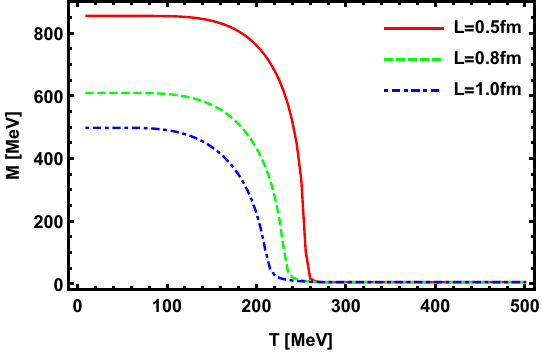}
	\caption{Quark mass as a function of temperature with only the lowest energy level is taken into account under different magnitude of magnetic field (top) as well as for different radius L for the cylinder (down), respectively.}
	\label{fig:catalysis}
\end{figure}

The quark mass as a function of the temperature under different magnitude of strong magnetic field is shown in the upper figure of Fig.\ref{fig:catalysis}. It is observed that both the quark mass and the corresponding critical temperature increase as the magnetic field increases, which is known as magnetic catalysis \cite{Gusynin:1994re,Miransky:2015ava}. Similarly, the catalysis of chiral condensate is also shown in the cylinder system as size decreases shown in the lower figure of Fig.\ref{fig:catalysis}. Small size plays similar role as the strong magnetic field, and they both enhance chiral symmetry breaking. The stronger the magnetic field and the smaller the size is, the more contribution from the zero-mode.

\begin{figure}	
	\centering
	\includegraphics[width=230pt]{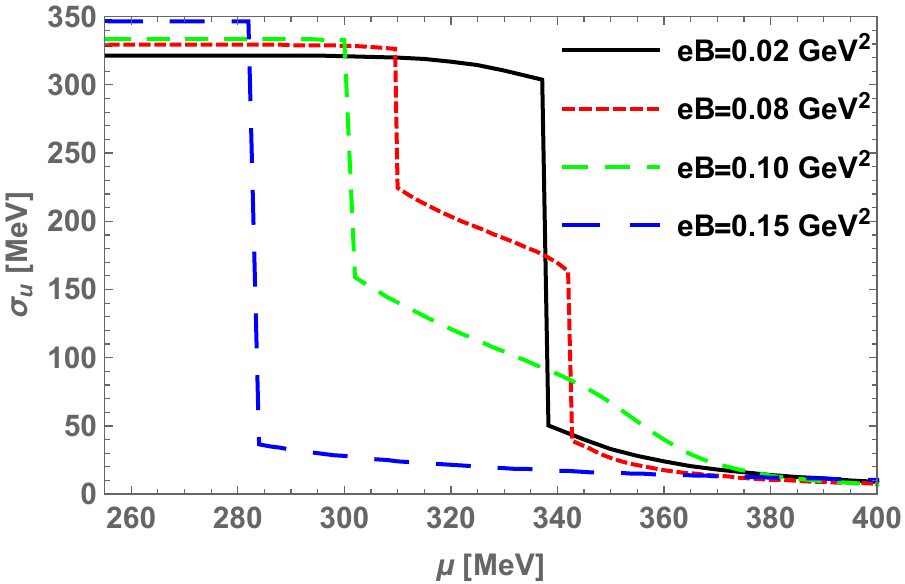}\\
	\includegraphics[width=230pt]{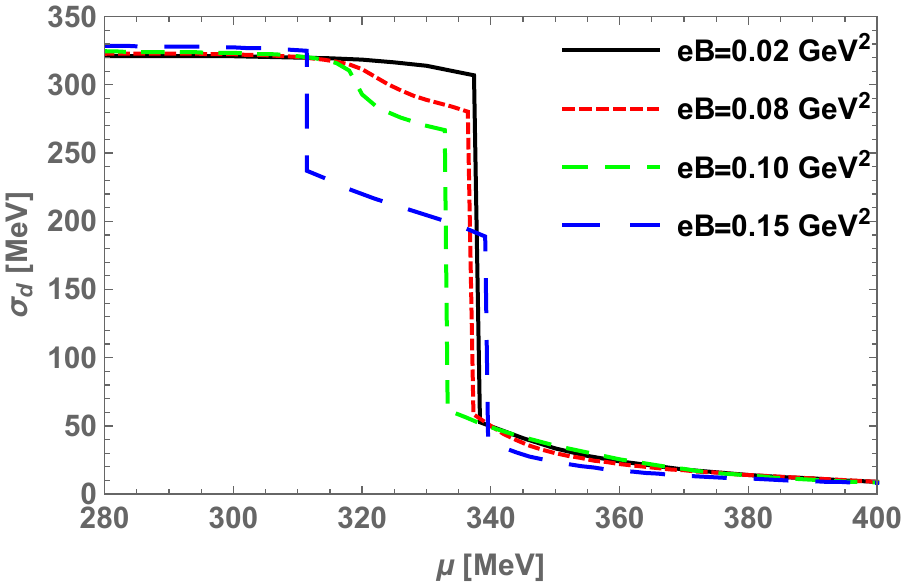}
	\caption{Chiral condensates for $u$ quark (upper figure) and $d$ quark (lower figure) as a function of the chemical potential for different magnitude of the magnetic field, respectively.}
\label{figure:quantized_pt_eB}
\end{figure}

Now we consider the cold dense quark matter under strong magnetic field. Because $u$ quark and $d$ quark carry different electric charges, therefore the fermi surfaces for $u$ and $d$ quarks are splitting, and their critical chemical potentials for chiral phase transitions are also different. The  chiral condensate at zero temperature for $u,d$ quarks at different magnitude of magnetic field are shown in Fig.\ref{figure:quantized_pt_eB}. As we expected, there are also quantized first-order chiral phase transition showing up for $u$ quark and $d$ quark, respectively. To our knowledge, the quantized first-order chiral phase transition phenomena under strong magnetic field has not yet been observed in other literatures except Ref. \cite{Costa:2013zca}.

\section{Summary and discussion}

Even though the finite size effect in QCD physics has attracted lots of interests for more than three decades, normally, both the periodic and the anti-periodic spatial boundary condition are applied for fermions. In this work, we find that if there is no other physical constraint, the ground state of quark matter favors the periodic spatial boundary condition, in which the zero-momentum mode is taken into account. In this stable small system, the catalysis of chiral symmetry breaking is observed with the decrease of the system size, while the pions excited from the droplet vacuum keep as pseudo Nambu-Goldstone bosons. The catalysis of chiral symmetry breaking and pseudo Nambu-Goldstone pions in small system are similar to those in quark matter under strong magnetic fields. The similarity between these two systems is understandable because the stronger the magnetic field the smaller the magnetic length of the charged particle will be.

For the periodic boundary condition, one can obtain the catalysis of chiral symmetry breaking in the vacuum at small size, which indicates the constituent quark becomes heavier and its wavelength becomes much smaller. We can imagine that the droplet-quark matter with periodic condition is a bag with dynamical massive quarks inside the bag surrounded by light pseudo Nambu-Goldstone pions' cloud.

Furthermore, it is found that the zero-momentum mode contribution brings significant change of the chiral apparent phase transition in a droplet of cold dense quark matter: the 1st-order chiral apparent phase transition becomes quantized, which is a brand-new quantized phenomena. We want to emphasize that this is the first time to observe the quantized 1st-order apparent phase transition in literatures. As we analyzed in this work, the quantized 1st-order apparent phase transition is induced by the quantized momentum and in this case the zero-momentum mode contribution becomes non-negligible. It is expected that such quantized 1st-order apparent phase transition is a common feature for system with quantized momentum. Therefore, we also observe such quantized 1st-order apparent phase transition phenomena in quark matter under strong magnetic field, and we expect such phenomena will also show up in some small systems in condensed matter.

At last, we expect that this new phenomena will largely affect the equation of state with small droplet quark matter inside neutron star thus affect the radius-mass relation of the neutron star. When consider the mixed phase of quark matter and nuclear matter, one has to solve the Wigner-Seitz cell structure, e.g., drop/bubble, rod/tube, and slab structure, the size of the Wigner-Seitz cell can be as small as several fms as shown in \cite{Wu:2017xaz}. For the small size of the Wigner-Seitz cell in neutron stars, the work of investigating how the zero-mode contribution will affect the equation of state and thus the neutron star properties is in progress \cite{XuKun-NeutronStar}. Also, it is noticed that in the future, it is worthy of performing calculations beyond mean-field approximation.

\begin{acknowledgments}
We thank valuable discussions with T.Hatsuda and R.Pisarski. This work is supported in part by the NSFC under Grant Nos. 11725523, 11735007, 11261130311 (CRC 110 by DFG and NSFC), Chinese Academy of Sciences under Grant No. XDPB09, and the start-up funding from University of Chinese Academy of Sciences(UCAS).

\end{acknowledgments}

\end{document}